\documentclass[preprintnumbers, pra, showpacs, floatfix,twocolumn,
preprintnumbers, letterpaper, superscriptaddress]{revtex4}
\usepackage{amsfonts}
\usepackage{amsmath}
\usepackage{amssymb,epsf}
\usepackage{latexsym}
\usepackage{graphicx,epsfig}
\usepackage{amssymb}
\usepackage{subfigure}
\usepackage[colorlinks=true,citecolor=blue,linkcolor=blue,urlcolor=black]{hyperref}
\usepackage{epstopdf}
\usepackage{color}

\def\be{\begin{equation}}
\def\ee{\end{equation}}
\def\ba{\begin{eqnarray}}
\def\ea{\end{eqnarray}}
\def\la{\langle}
\def\ra{\rangle}

\begin{document}

\title{Dual correspondence between classical spin models and quantum CSS states}
\author{Mohammad Hossein Zarei}
\email{mzarei92@shirazu.ac.ir}
\affiliation{Physics Department, College of Sciences, Shiraz University, Shiraz 71454, Iran}
\author{Afshin Montakhab}
\email{montakhab@shirazu.ac.ir}
\affiliation{Physics Department, College of Sciences, Shiraz University, Shiraz 71454,
Iran}

\begin{abstract}

The correspondence between classical spin models and quantum
states has attracted much attention in recent years. However, it
remains an open problem as to which specific spin model a given
(well-known) quantum state maps to. Here, we provide such an
explicit correspondence for an important class of quantum states
where a duality relation is proved between classical spin models
and quantum Calderbank-Shor-Steane (CSS) states. In particular, we
employ graph-theoretic methods to prove that the partition
function of a classical spin model on a hypergraph $H$ is equal to
the inner product of a product state with a quantum CSS state on a
dual hypergraph $\tilde{H}$. We next use this dual correspondence
to prove that the critical behavior of the classical system
corresponds to a relative stability of the corresponding CSS
state to bit-flip (or phase-flip) noise, thus called
\emph{critical stability}. We finally conjecture that such
critical stability is related to the topological order in quantum
CSS states, thus providing a possible practical characterization
of such states.

\end{abstract}


\maketitle

\section{Introduction} Quantum entangled states play an
important role in quantum computation and quantum communication.
Of particular interest are error correcting codes such as CSS
states \cite{Calderbank1996,stean,gottesman,cssd} which are used for
protecting information against decoherence. Because of their high
entanglement properties, they have also found many applications in
quantum information protocols such as encryption
\cite{Hillery1999} and measurement-based quantum computation
(MBQC) \cite{Raussendorf2001,nature}. Furthermore, an important
set of such states called topological CSS states, including
Kitaev's toric code states (TC) \cite{Kitaev2003} and color code states (CC)
\cite{bombin2006}, have found important applications as robust
memory for topological quantum computation. Such a robustness
against local perturbations is a result of topological order which
is a new phase of matter that is not described by symmetry
breaking theory \cite{Wen1990,Brown2016,Brown2016r}.

On the other hand, statistical physics is a well-established field
with many interdisciplinary applications in recent decades, in
particular with regards to criticality and phase transitions
\cite{biology, economics, chaos}. Accordingly, connections between
quantum information theory and classical statistical mechanics has
attracted much attention
recently\cite{Geraci2008,Lidar1997,Geraci2010,Somma2007,Verstraete2006,mont2010},
which has led to cross-fertilization. Specifically in an important
paper \cite{Nest2007}, it was shown that a partition function of a
classical spin model can be written as an inner product of an
entangled state and a product state. Such a mapping has found
many applications in both quantum information theory and
statistical mechanics
\cite{Cuevas2009,Cuevas2011,xu,Vahid2012, yahya,science}. For example, it
is shown that computational power of quantum states for
measurement-based quantum computation is related to computational
complexity of the corresponding classical spin models. Using such
a relation, it is shown that some specific classical spin models,
corresponding to quantum stabilizer states which are universal
resources for MBQC, are complete \cite{Nest2008,Vahidb} in a sense
that their partition function is equivalent to partition functions
of any other classical spin models. Furthermore, a similar
correspondence between some classical spin models with specific
topological code states has been exploited in order to study the
power of a few topological states as a source for MBQC
\cite{Bravyi2007,Bombin2008}.

Although the above applications of the correspondence between the
partition function of classical spin models and quantum entangled
states has only been considered for a few specific models, one
might expect that a wider class of classical spin models could
have specific quantum state mapping. Accordingly, finding an
explicit correspondence between specific spin models and various
quantum states is an important and challenging task. In
particular, one might like to know what are the classical spin
models corresponding to well-known quantum states. To this end,
one needs to find a common tool which  can be employed as a mapping
mechanism for both classical spin models as well as quantum
entangled states. In this paper, we use hypergraphs as such a tool
to find an explicit mapping to establish a duality correspondence
between the partition function of classical spin models and
quantum CSS states. In particular, we show that the classical spin
models map to a hypergraph which corresponds to a quantum CSS
state on the corresponding dual hypergraph. Our results are very
general and can in principle be applied to a wide class of models,
as in spatial dimensions higher than the typical 1D or 2D systems.
Furthermore, in order to show that hypergraph is a useful and
practical tool and provide some specific examples, we show the
details  of such mapping for TC on arbitrary graphs as well as CC
on D-colexes.

In addition to the possibility of the above-mentioned
applications, one might also think of other possible insights. For
example, we show that one can use the well-known properties of the
classical systems in order to obtain important new information
about the corresponding quantum CSS states.  In particular, we use
the non-analytic properties of classical partition functions at
the critical point of their second order phase transitions
($T=T_{cr}$) in order to draw conclusions about the corresponding
CSS states, which become relatively stable at the particular value
of noise, precisely related to $T_{cr}$. We therefore call this
new concept \emph{critical stability}. This type of
stability is distinctly different from the more common concept of robustness and the accuracy threshold of
CSS codes associated with classical spin glasses which have been
previously studied \cite{Dennis2002,q1,q2,q3,q4,Katzgraber2009}.
We further conjecture that critical stability as an intrinsic
physical property can be used to ascertain topological order in
quantum CSS states.

This paper consists of the following sections: in Sec.(\ref{s1})
we provide a basic review on hypergraphs which we use to establish
our duality mapping. In Sec.(\ref{s2}) we establish our duality
mapping where we prove that the partition function of a classical
spin model on a hypergraph corresponds to CSS state on a dual
hypergraph. We next provide some specific examples of such duality
for the well known TC as well as the CC in Sec.(\ref{s3}). Next as an interesting application of
the duality, we prove that the critical phase transition of the
classical spin model corresponds to the stability of the CSS
state \textit{at a specific} noise probability, thus referred to
as critical stability. We end in Sec.(\ref{s4}) by providing some
concluding remarks.

\section{Basics of hypergraphs}\label{s1}
\begin{figure}[t]
\centering
\includegraphics[width=8cm,height=3cm,angle=0]{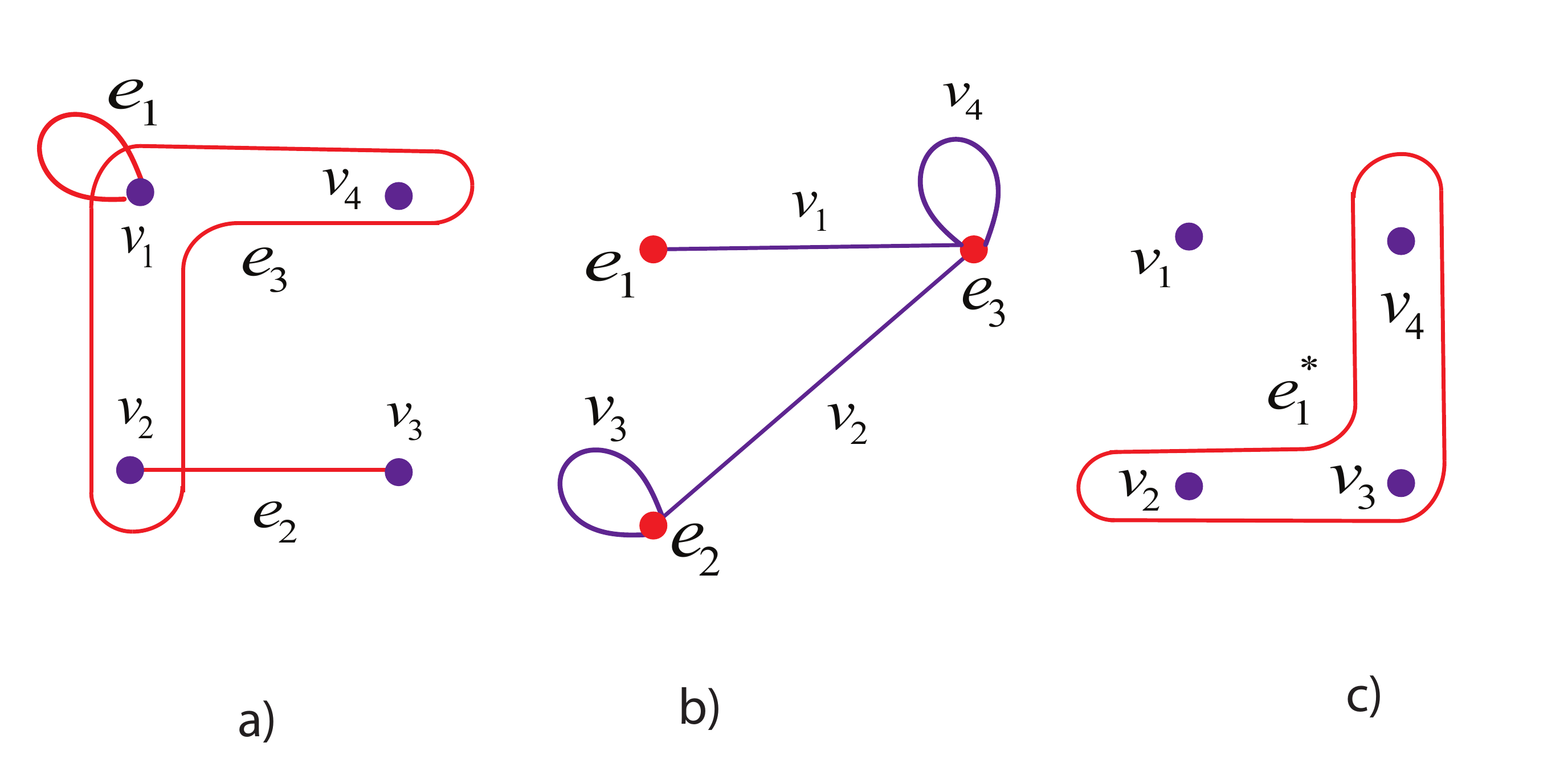}
\caption{(Color online) (a) A simple hypergraph where we use purple (dark) color for vertices and yellow (light) color for hyperedges. We use a loop for an edge containing only one vertex, a link for
edges containing two vertices and closed curves for edges
containing more than two vertices. (b) Dual hypergraph where we use yellow (light) color for vertices and purple (dark) color for hyperedges  and (c)
orthogonal hypergraph of part (a).} \label{hyp}
\end{figure}

A hypergraph $H$ is an
extension of an ordinary graph where each edge of the hypergraph
can involve arbitrary number of vertices. In this way, each
hypergraph is characterized by two sets of vertices $ V=\{v_1 ,
v_2 , ...,v_K \} $ and (hyper)edges $E=\{e_1 , e_2 , ...,e_N \}$
and is denoted by $H=(V,E)$. As an example in Fig.\ref{hyp}(a), a
hypergraph of four vertices $v_1, v_2 , v_3 , v_4$ has been shown
where edge  $e_{1} =\{v_1 \} $, edge  $e_{2} =\{v_2 , v_3\} $,
and $e_{3} =\{v_1 , v_2 , v_4\} $, which is denoted by a closed
curve. Degree of each edge $e_{m}$ is called cardinality and is
equal to the number of vertices that are involved by $e_{m}$,
denoted by $|e_{m}|$. In Fig.\ref{hyp}(a), degree of edges
$e_{1}$, $e_{2}$, $e_{3}$ are equal to 1, 2, and 3, respectively. The
dual of a hypergraph $H(V,E)$ is a hypergraph
$\tilde{H}=(\tilde{V},\tilde{E})$ where $\tilde{V}=\{\tilde{v}_1 ,
\tilde{v}_2 ,...,\tilde{v}_N \}$ and $\tilde{E}=\{\tilde{e}_1 ,
\tilde{e}_2 , ..., \tilde{e}_K\}$ where $\tilde{e}_i=\{\tilde{v}_m
| v_i \in e_m~in~ H\}$, i.e. duality interchanges vertices and
edges \cite{11}. For example, in Fig.\ref{hyp}(b), we show the
dual hypergraph of part (a).

Consider a hypergraph with $|V|=K$ vertices and $|E|=N$ edges.
Related to each edge $e_{m}$, we consider a binary vector, which
is called edge vector, with $K$ components which are denoted by
$a_{m}^{j}$ and $j=\{1,2,3,...,K\}$ where $a_{m}^{j}=1$ if $v_{j}
\in e_{m}$ and $a_{m}^{j}=0$, otherwise. For example, for
hypergraph in Fig.\ref{hyp}(a), the corresponding binary vectors
to three edges are $e_{1} =(1,0,0,0)$, $e_{2} =(0,1,1,0)$ and
$e_{3} =(1,1,0,1)$ . In this way, we will have $N$ binary vectors
corresponding to $N$ edges of the hypergraph. Furthermore, an edge
vector is called dependent if it can be written as a superposition
of other edge vectors of the hypergraph. The set of all
independent edges is an independent set which is denoted by
$\mathcal{I}$, where $| \mathcal{I} | \leq |V|$. We can also
define an orthogonality relation between edges where two edges
$e_{m}$ and $e_{n}$ are called orthogonal if and only if their
corresponding binary vectors are orthogonal, i.e. $e_{m}.e_{n}=0$ where the symbol ``." refers to the inner product,
in binary representation. Consider a hypergraph $H=(V,E)$ with an
independent set of edges, $\mathcal{I}$, orthogonal hypergraph of
$H$ is a hypergraph $H^{*}=(V^{*},E^{*})$ that has the same
vertices as $H$, $V^{*}=V$, but has $K-|\mathcal{I}|$ distinct and
independent edges that are orthogonal to all edges of $H$, see
Fig.\ref{hyp}(c).

 One can show that, for any hypergraph $H$, an orthogonal hypergraph $H^{*}$
always exists. To this end, consider a
hypergraph $H=(V,E,\mathcal{I})$ where $\mathcal{I}$ is the independent set of the
hyperedges. If $|V|=K$ and $|\mathcal{I}|=M$, it is clear that the number of
independent edges is smaller than the number of vertices, i.e.
$M<K$. We denote the independent hyperedges by $e_1 , e_2 , ...,
e_M$ with a binary vector representation for each one of them.

We want to find a new hypergraph $H^{*}$ that is orthogonal to
$H$. To this end, we need to find all hyperedges denoted by
$e^{*}$ which are orthogonal to all $e_i$. The binary vector of
$e^{*}$ has $K$ components denoted by $a_{1}^{*}, a_{2}^{*},...,
a_{K}^{*}$. Clearly, $e^{*}$ will be orthogonal to all edges of
the $H$ if and only if, for $i=1,2,...,M$, we have $e^{*}.e_i =0$.
Such a relation for all $i$ is equivalent to $M$ independent
equations on $K$ binary variables $a_{1}^{*}, a_{2}^{*},...,
a_{K}^{*}$. The number of independent solutions for such a set of
equations is equal to $K-M$. Since each independent solution is
equal to a independent hyperedge of $H^{*}$, we will find an orthogonal hypergraph $H^{*}$ for each original
hypergraph $H$ with a distinct independent set of hyperedges

\section{Duality mapping between classical spin models and quantum CSS states}\label{s2}
 One can use hypergraphs in order to define the most general form of spin
Hamiltonian:
\begin{equation}
\mathbb{H}=\sum_{i}J_i s_i +\sum_{i,j} J_{ij}s_i s_j +\sum_{i,j,k} J_{ijk}s_i s_j s_k +...
\end{equation}
where $s_i=\{\pm1\}$, $J_i$ referes to local magnetic field, $J_{ij}$ and $J_{ijk}$ refers to two-body and three-body coupling constants, and $+...$ refers to other many-body interactions. Such form of Hamiltonian can also include
d-level spins as is shown in \cite{Cuevas2009}. To map the spin
model to a hypergraph $H(V,E)$, spins are represented by the
vertices, and corresponding to each interaction term, we define an
edge which includes all spins belonging to the interaction term.
Therefore, we can re-write the above Hamiltonian in a compact
form:
\begin{equation}\label{a1}
\mathbb{H}=\sum_{m|  e_{m} \in E}J_m \prod_{i|v_i \in e_m}s_i .
\end{equation}

We next map a quantum
CSS state to a hypergraph, see also \cite{zareiprb}. To this end, we use an idea which has recently been used for some quantum entangled states in \cite{qhyp1, qhyp2}. A quantum CSS state
on $K$ qubits is a stabilizer state that is stabilized by $X$-type
and $Z$-type operators belonging to Pauli group on $K$ qubits. To
encode the structure of quantum CSS states, consider a hypergraph
$H=(V,E,\mathcal{I})$, with $|V|=K$, $|E|=N$ and
$|\mathcal{I}|=M$. We insert $K$ qubits in all vertices of the
hypergraph. Then we define an $X$-type operator corresponding to
each independent edge of $H$, and also define a $Z$-type operator
corresponding to each edge of the $H^{*}=(V,E^{*})$. We denote
such operators by $A_{m}$ and $B_{n}$, respectively, where $e_{m}
\in \mathcal{I}$ and $e^{*}_n \in E^*$, and are given by:
\begin{equation}
A_{m} =\prod_{i| v_i \in e_{m}}X_i~~,~~B_{n} =\prod_{i| v_i \in e^{*}_{n}}Z_i .
\end{equation}
Clearly,  when two edges are orthogonal to each other it is
necessary that the number of shared vertices of those edges is an
even number, therefore each $X$-type operator commutes with other $Z$-type operators, i.e. $[A_{m} , B_{n}]=0$. In this way,
corresponding to $H$, we will have $M$, $X$-type stabilizers and
$K-M$, $Z$-type stabilizers which are generators of a quantum CSS
state in the following form: \begin{equation}\label{q1}
|CSS_H\ra=\frac{1}{2^{\frac{M}{2}}}\prod_{m| e_{m} \in
\mathcal{I}}(1+A_{m} )|0\ra ^{\otimes K}
\end{equation}
where $|0\ra$ is the positive eigenstate of $Z$ and $\frac{1}{2^{\frac{M}{2}}}$is the normalization factor. Since $A_{m}
(1+A_{m})=(1+A_{m})$ and $[A_{m} , B_{n}]=0$, one can easily check
that the above state is stabilized by $A_m$ and $B_n$.

Furthermore, it is also possible to represent  Eq.(\ref{q1}) with
$Z$-type operators $B_{n}$ in the following form:
\begin{equation}\label{q2}
|CSS_H\ra=\frac{1}{2^{\frac{K-M}{2}}}\prod_{n| e^{*}_{n} \in E^{*}}(1+B_{n} )|+\ra ^{\otimes K}
\end{equation}
where $|+\ra$ is the positive eigenstate of $X$. Therefore, we can
construct a CSS state corresponding to a hypergraph with a
distinct independent set.

 We are now ready to present our main result. Consider a
classical spin model on a hypergraph $H=(V,E)$ given by
Eq.(\ref{a1}). The partition function is given by:
\begin{equation}\label{main}
\mathcal{Z}=\sum_{\{s_i\}}exp[-\beta \sum_{m|e_{m}\in E}J_m \prod_{i| v_i \in
e_{m}}s_i]
\end{equation}
where $\beta$ is the reciprocal temperature with the Boltzman constant $k_B$ sets equal to one. We can perform a simple
change of variable. To this end, corresponding to each edge $e_{m}$, we
define a new spin variable $S_{m}$ which is related to all the spins
$s_i$ belonging to the edge $e_{m}$ in the form of $S_m =\prod_{i| v_i \in
e_{m}}s_i$, and replace these new \emph{edge variables} in the above
relation. It is clear that these new variables are not necessarily
independent variables.

\begin{figure}[t]
\centering
\includegraphics[width=8cm,height=3cm,angle=0]{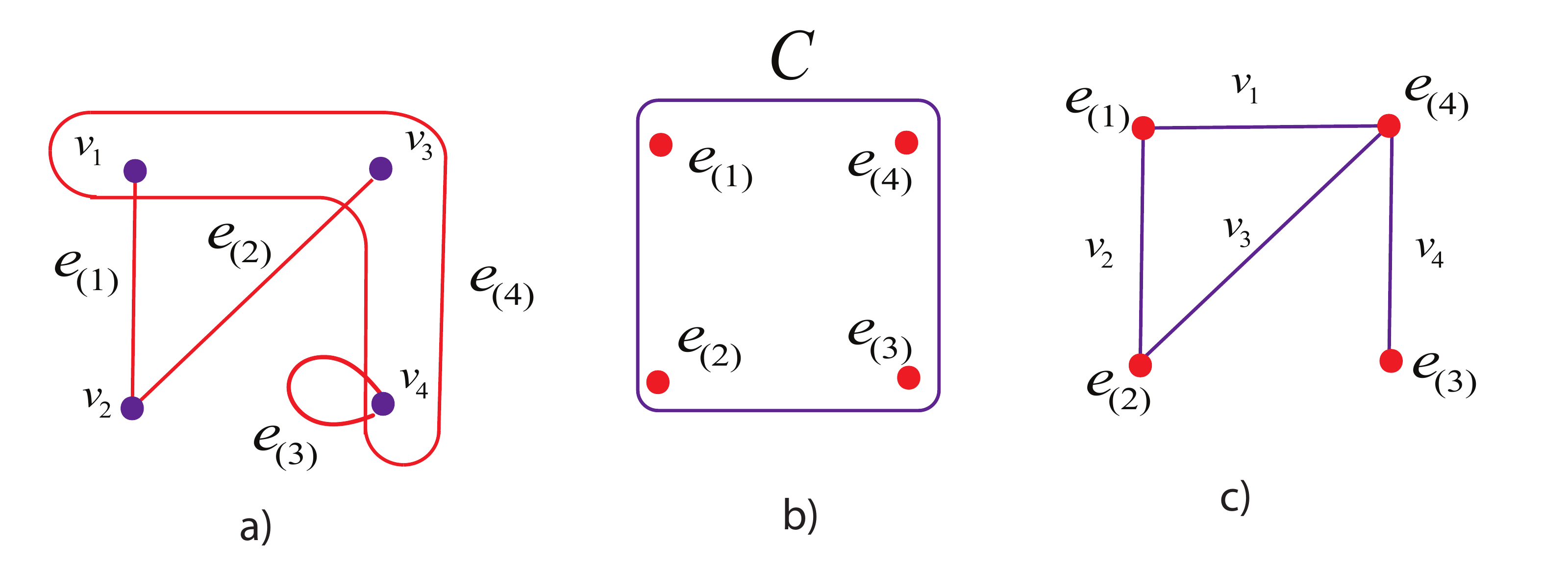}
\caption{(Color online) (a) A hypergraph corresponding to a simple
spin model. Spin variables correspond to vertices and each edge
refers to an interaction term. (b) Corresponding to each edge of
the original hypergraph, a new vertex is assigned  with a new spin
variables $S_{(m)}$. A constraint on these variables is denoted by
a closed curve. This new hypergraph is denoted $H_C$. (c) The dual
of the original hypergraph  is orthogonal to $H_C$.} \label{exa}
\end{figure}

To better understand this change of variable, consider a spin
model on a hypergraph as shown in Fig.\ref{exa}(a), with four
vertices $v_1 , v_2 , v_3 , v_4$ and four edges $e_{1} =\{v_1 ,
v_2 \}$, $e_{2} =\{v_2 , v_3 \}$, $e_{3} =\{v_4 \}$ and $e_{4}
=\{v_1 , v_3 , v_4 \}$. The classical spin model corresponding to
this hypergraph will be given by: $\mathbb{H}=-J_{(1)}s_1
s_2-J_{(2)}s_2s_3-J_{(3)} s_4-J_{(4)} s_1 s_3 s_4$ where the
variable $s_i$  corresponds to the vertex $v_i$ and $J_{m}$ refers
to coupling constant of the edge $e_{m}$. We now replace the
original spin variables $s_i$ by the edge variables $S_{m}$:
$S_{(1)}=s_1s_2$, $S_{(2)}=s_2s_3$, $S_{(3)}=s_4 $ and
$S_{(4)}=s_1s_3s_4 $. By such a definition for edge variables, it
is clear that there is a constraint given by
$S_{(1)}S_{(2)}S_{(3)}S_{(4)}=1$.  In Fig.\ref{exa}(b), we have
shown this constraint by a closed curve around the four new spin
variables corresponding to edges $e_m$. It is necessary to apply
this constraint in the expression for the partition function. We
thus use a Kronecker (Leopold Kronecker) delta as
$\delta(S_{(1)}S_{(2)}S_{(3)}S_{(4)},1)$ and the partition
function can be written as:
$\mathcal{Z}=\sum_{\{S_{m}\}}\delta(S_{(1)}S_{(2)}S_{(3)}S_{(4)},1)
\times
exp[\beta(J_{(1)}S_{(1)}+J_{(2)}S_{(2)}+J_{(3)}S_{(3)}+J_{(4)}S_{(4)})]$.
It is clear that the above example can be extended to any spin
model and we can denote the set of all edges $e_m$ belonging to a
constraint $C$ where such a constraint can be written as
$\prod_{m| e_{m}\in C}S_{m} =1$. However, it is very important to
find a simple interpretation for all constraints in a general way.
To this end, we define a new hypergraph $H_C$, whose vertices are
the variables $S_{m}$. We consider each constraint $C$ as an edge
of hypergraph $H_C$. Since each vertex of $H_C$ is related to an
edge of $H$, we denote those vertices by $\tilde{v}_m$ in analogy to
vertices of the dual hypergraph $\tilde{H}$. Therefore, $H_C$ is a
hypergraph in the dual space with a set of vertices corresponding
to spin variables $S_{m}$, which we denote by $\tilde{v}_m$, and a
set of edges corresponding to constraints $C$. A simple lemma can
now be proven in order to interpret $H_C$.

\textit{Lemma:} $H_C$ is equal to the orthogonal hypergraph of
dual hypergraph $\tilde{H}$, i.e. $H_C = \tilde{H}^{*}$.
\textit{Proof:} We prove this lemma in two steps. In the first
step, since $S_{m}=\prod_{i| v_i \in e_{m}}s_i$, it is clear that
each constraint in the form of $\prod_{m| e_{m}\in C} S_{m}=1$
will hold true if and only if each vertex of $H$ is a member of an
even number of edges belonging to constraint $C$. Now consider the
above conclusion in a dual space where vertices of the hypergraph
$H$ are edges of $\tilde{H}$ which are denoted by $\tilde{e}_i$,
and constraints $C$ are denoted by $\tilde{e}_C$, the edges of
$H_C$. Therefore, in a dual space, $|\tilde{e}_i \bigcap
\tilde{e}_C|$ is an even number. In the second step, consider
binary vectors corresponding to edges $\tilde{e}_i$ and
$\tilde{e}_C$ in the dual space. Since $|\tilde{e}_i \bigcap
\tilde{e}_C|$ is an even number, it is simple to show that
$\tilde{e}_i . \tilde{e}_C=0$.  Therefore, the binary vectors
corresponding to constraints in $H_C$ are orthogonal to binary
vectors corresponding to edges of the dual hypergraph $\tilde{H}$.
In other words, $H_C$ is equal to the orthogonal hypergraph of the
dual hypergraph $\tilde{H}$, and thus, the lemma is proved. As an
example, in Fig.\ref{exa}(c), we show dual hypergraph of
Fig.\ref{exa}(a). It is simple to check that $H_C$ is orthogonal
to $\tilde{H}$.

To recap, $H_C =\tilde{H}^{*}$ and each constraint $C$ on spin
variables $S_m$ ($\prod_{m|e_{m}\in C}S_{m} =1$) can be written as
$\prod_{m| \tilde{v}_{m} \in \tilde{e}^{*}}S_m =1$ where
$\tilde{v}_m$ and $\tilde{e}^{*}$ refer to a vertex and edge of
$\tilde{H}^{*}$, respectively. We are ready to come back to
Eq.(\ref{main}) for the partition function of classical spin
models which now finds the form:
\begin{equation}\label{sa}
\mathcal{Z}=\sum_{\{S_{m}\}}e^{(-\beta \sum_{m| \tilde{v}_m \in
\tilde{V}}J_{m} S_{m})}\prod_{\tilde{e}^{*} \in
\tilde{E}^{*}}\delta(\prod_{m| \tilde{v}_m \in \tilde{e}^{*}}S_m
,1),
\end{equation}
where $\tilde{V}$ and $\tilde{E}^{*}$ are sets of vertices and edges of $\tilde{H}^{*}$, respectively.
In the next step, we show that the above form of the partition
function can be written in a quantum language. To this end, we use
a simple identity for the Kronecker delta in the form of $\delta(\prod_{m| \tilde{v}_m \in
\tilde{e}^{*}}S_m,1)= \frac{(1+\prod_{m| \tilde{v}_m \in
\tilde{e}^{*}}S_m )}{2}$ and rewrite the partition
function as
\begin{equation}\label{sb}
\mathcal{Z}=\sum_{\{S_{m}\}}e^{(-\beta \sum_{m| \tilde{v}_m \in
\tilde{V}}J_{m} S_{m})}\prod_{\tilde{e}^{*} \in
\tilde{E}^{*}}\frac{(1+\prod_{m| \tilde{v}_m \in
\tilde{e}^{*}}S_m )}{2} .
\end{equation}
Since each spin variable has a value of $1$ or $-1$, it is simple
to show that a summation $\sum_{\{S_m\}}F(S_{m})$, where $F$ is an
arbitrary function, can be written as $ \sum_{\{S_m
\}}F(S_{m})=2\la+|F(Z)|+\ra$.  Therefore,
$$ \mathcal{Z}=2^{N}  ~^{N\otimes}\la+|e^{(-\beta\sum_{m| \tilde{v}_m \in
\tilde{V}}J_{m}Z_{m})}$$
\begin{equation}\label{sb1}
.\prod_{\tilde{e}^{*} \in
\tilde{E}^{*}}\frac{(1+\prod_{m| \tilde{v}_m \in
\tilde{e}^{*}}Z_m
)}{2}|+\ra ^{\otimes N}
\end{equation}
where $|+\ra$ refers to positive eigenstate of the $X$ operator and $N$ is the number of vertices of the hypergraph
$\tilde{H}^{*}$. If we consider the number of independent edges of the
$\tilde{H}$ to be equal to $M$, we can write the above as:
\begin{equation}\label{sc}
\mathcal{Z}=2^{M}\la \alpha | Q\ra
\end{equation}
where $|\alpha \ra $ is a product state given by $ \prod_{m|
\tilde{v}_m \in \tilde{V}}exp(-\beta J_{m}Z_{m})|+\ra^{\otimes N}$
and $ |Q\ra = \prod_{\tilde{e}^{*} \in \tilde{E}^{*}}(1+\prod_{m|
\tilde{v}_m \in \tilde{e}^{*}}Z_m )|+\ra ^{\otimes N} $  is an unnormalized
quantum state. Comparing with Eq.(\ref{q1}, \ref{q2}), we conclude
that $|Q\ra$ is equal to a quantum CSS state on dual hypergraph
$\tilde{H}$ up to a normalization factor: $$ |Q\ra = \prod_{\tilde{e}^{*} \in
\tilde{E}^{*}}(1+\prod_{m| \tilde{v}_m \in \tilde{e}^{*}}Z_m
)|+\ra ^{\otimes N}$$
\begin{equation}
=2^{\frac{N}{2}-M} \prod_{\tilde{e}\in \tilde{\mathcal{I}}}(1+\prod_{m| \tilde{v}_m \in \tilde{e} }X_m )|0\ra ^{\otimes N}= 2^{\frac{N-M}{2}}| CSS_{\tilde{H}}\ra ,
\end{equation}
where $\tilde{\mathcal{I}}$ is the independent set of $\tilde{H}$. We have therefore proven that the partition function of a
classical spin model on $H$ corresponds to a quantum CSS state on
$\tilde{H}$ in the following form:

\begin{equation}\label{sd}
\mathcal{Z}_{H}=2^{\frac{N+M}{2}}\la \alpha | CSS_{\tilde{H}}\ra.
\end{equation}

\section{examples}\label{s3}
Using the duality relation Eq.(\ref{sd}), one can find specific
classical spin models corresponding to well-known quantum CSS
states. In this section, we show this by considering two important
classes of quantum CSS states, i.e. the TC and the
CC.

\subsection{Kitaev's toric code state and Ising model on arbitrary graphs}
First, we consider the duality mapping for TC
on an arbitrary graph which is a CSS quantum state with
topological order \cite{Kitaev2003}. We show that TC on an
arbitrary graph (lattice) corresponds to the Ising model on the
same graph (lattice). To this end, consider an arbitrary graph $G$
where qubits live on the edges of $G$. Corresponding to any
vertices of the graph one defines $X$-type operators in the
following form:
\begin{equation}
A_v =\prod_{i \in v} X_i
\end{equation}
where $i\in v$ refers to qubits that live on the edges of vertex
$v$. For TC the $Z$-type operators that commute with $A_v$ can
easily be found, where corresponding to each plaquette of the
graph, an operator  $B_p =\prod_{i\in \partial p}Z_i$ is defined,
where $i \in \partial p$ refers to qubits belonging to the
boundary of the plaquette $p$ for a square lattice, see
Fig.\ref{kit}(a). In this way, the TC is a stabilizer state of
both $X$ and $Z$-type operators in the following form:
\begin{equation}
|K_G \ra =\prod_{v}(1+A_v )|0\ra^{\otimes N}.
\end{equation}
Although the above state has been defined on a graph, it is simple
to present a hypergraph representation for it. To this end, we
consider all qubits of the TC as vertices of a hypergraph. Then,
corresponding to each vertex of graph $G$, we define a hyperedge
of the hypergraph $H$ that involves all qubits connecting to that
vertex, see Fig.\ref{kit}(b). In this way $X$-type operator $A_v$
for the TC can be denoted by $A_e$ where $e$ refers to a hyperedge
of $H$.

\begin{figure}[t]
\centering
\includegraphics[width=7cm,height=4.5cm,angle=0]{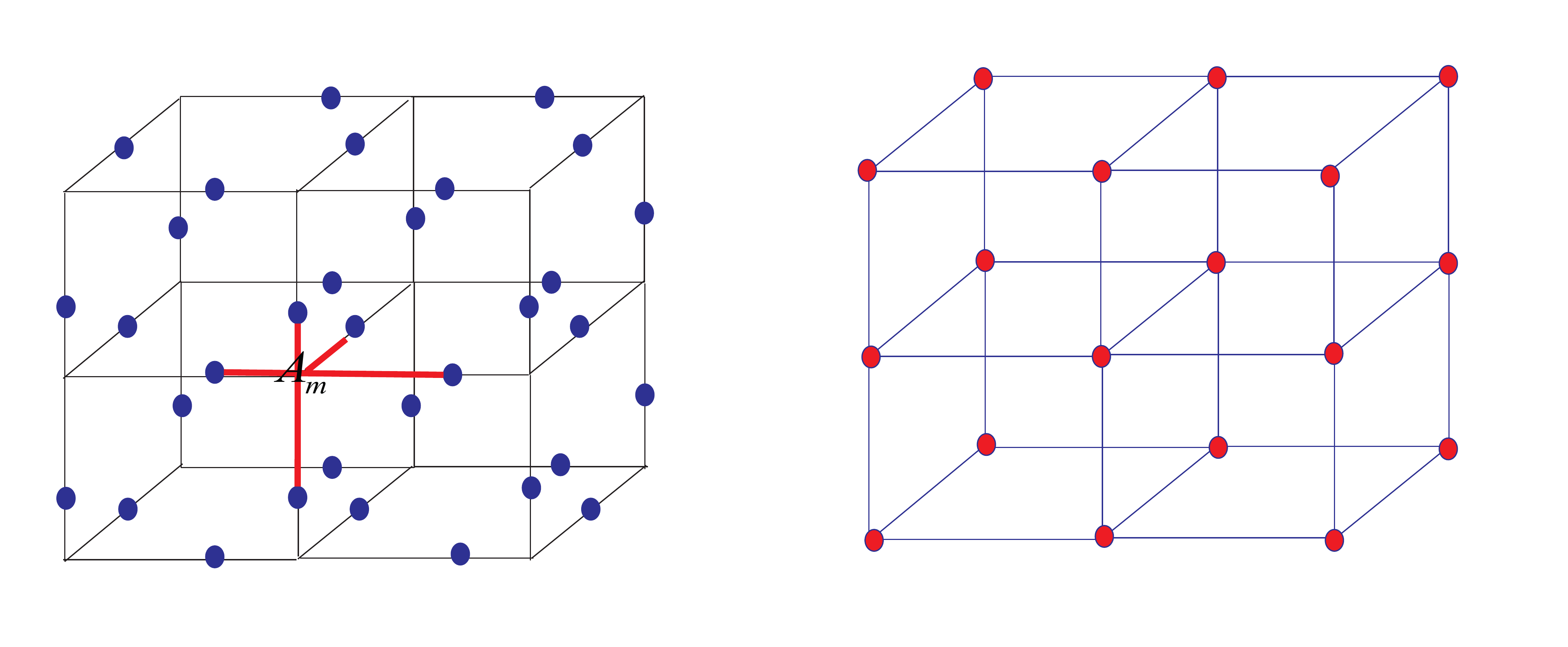}
\caption{(Color online) left: TC on a 3D lattice, each node of the
lattice corresponds to a hyperedge of a hypergraph which is
denoted by yellow (light) color. right: In the dual space, a vertex of the
$\tilde{H}$ denoted by yellow (light) circle corresponds to each edge of the
$H$. Each edge of the $\tilde{H}$ denoted by a purple (dark) link involves
two vertices. } \label{3Dtoric}
\end{figure}

According to the our duality mapping, a TC on a hypergraph $H$ is
related to a spin model on the dual hypergraph $\tilde{H}$. In
order to find the dual hypergraph relating to the TC on a graph
$G$, we insert spin variables on vertices of the $G$ corresponding
to each edge of $H$, see Fig.\ref{kit}(c). In this way, the spin
variables are considered as vertices of the $\tilde{H}$.
Furthermore, each vertex of $H$ is equal to a hyperedge of
$\tilde{H}$ and since each vertex of $H$ is a member of two
neighboring hyperedges of $H$, each hyperedge of $\tilde{H}$
should involve two spin variables, see Fig.\ref{kit}(c).
Therefore,  the hypergraph $\tilde{H}$ is an ordinary graph that
is exactly the same as the original graph $G$ and the
corresponding spin models will be an Ising model on the graph $G$.
In Fig.\ref{3Dtoric}, we show another example of a TC on a 3D
square lattice where the same argument as above holds.
Accordingly, the following relation holds between the partition
function of Ising models on an arbitrary graph $G$ and the TC
state $|K_{G}\ra$:
\begin{figure}[t]
\centering
\includegraphics[width=7cm,height=4.5cm,angle=0]{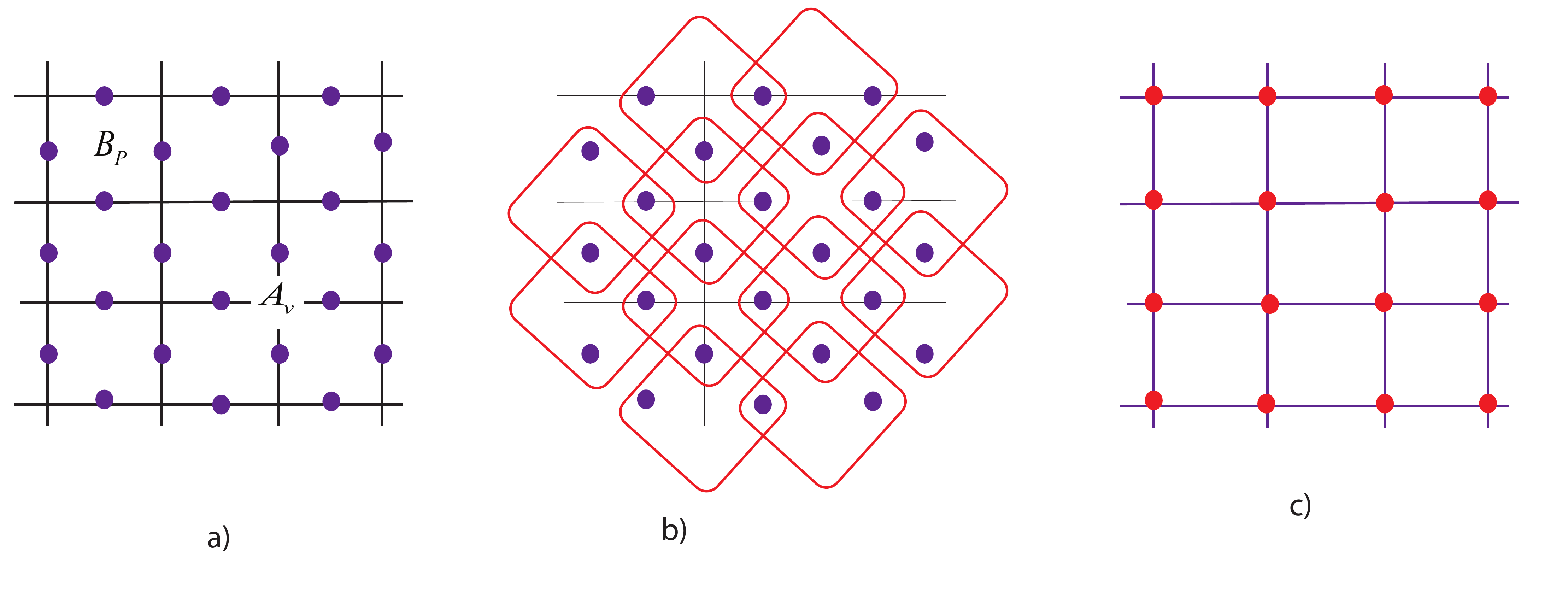}
\caption{(Color online) a) Plaquette and vertex operators of the
TC are shown on a simple square lattice. b)
Corresponding to each vertex of the original graph we consider an
edge of a hypergraph denoted by yellow (light) color where the TC state can be re-defined as a
quantum CSS state on that hypergraph. c) Since neighboring edges
of the hypergraph have only one common vertex, dual of the
hypergraph is the same original graph.} \label{kit}
\end{figure}
\begin{equation}\label{td0}
\mathcal{Z}_{Ising, G}=\la \alpha |K_{G}\ra .
\end{equation}
\subsection{Color code state on D-colexes and spin model on D-dimensional simplicial lattice}
Another set of quantum CSS states with topological order are CC. They can be defined on $(D+1)$-valent lattices
with $(D+1)$-colorable edges in D-dimensional manifold which is
technically called D-colexes \cite{delgad}. We next use
the duality mapping to show that inner product of a product state
with a CC is equal to the partition function of a spin model on a
simplicial lattice.

In order to show the above result, we present the main idea of CC
on two simple easy-to-imagine lattices, and then generalize to
higher dimensions. In two dimensions, a 2-colex is a trivalent
lattice where the edges can be colored by three different colors
such that any two neighboring edges do not have the same color. In
Fig.\ref{col}(a), we show an example of such a structure.
Corresponding to each plaquette of a 2D trivalent lattice, we
define two $X$-type and $Z$-type operators in the following form:
\begin{figure}[t]
\centering
\includegraphics[width=7cm,height=4.5cm,angle=0]{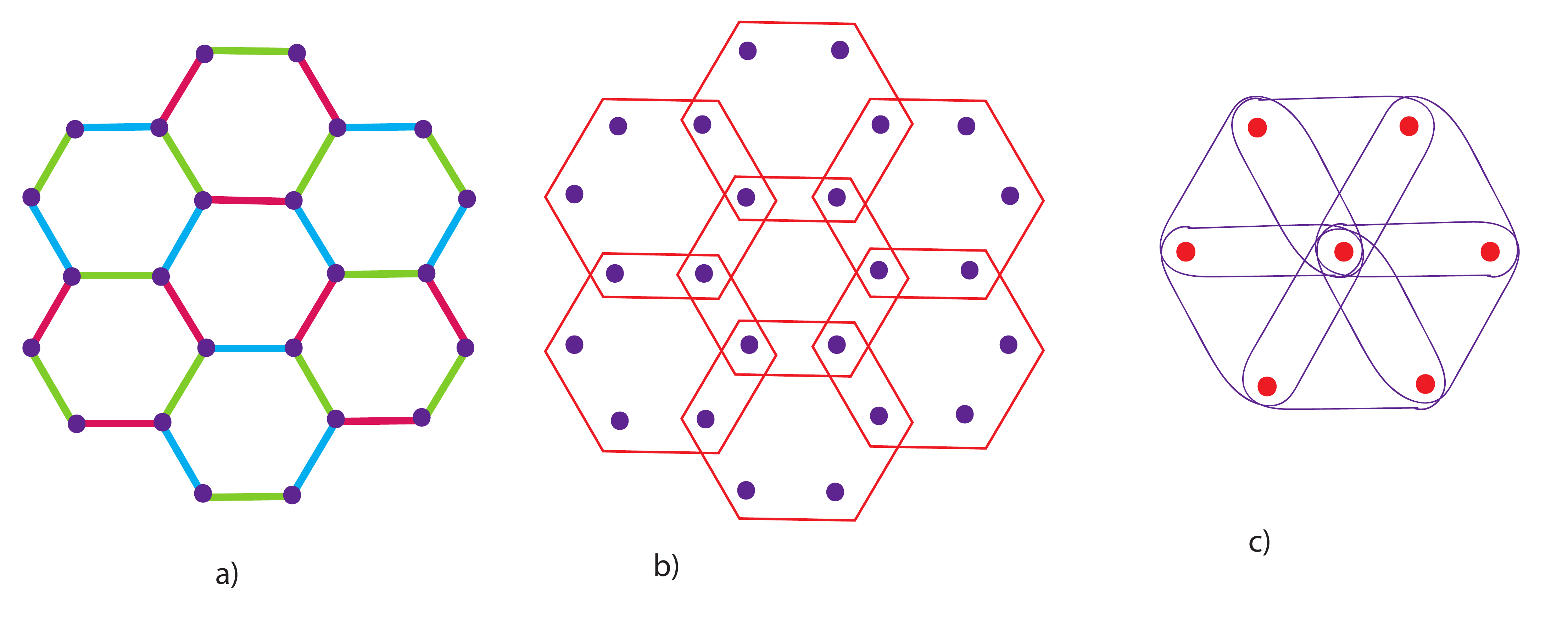}
\caption{(Color online) a) A hexagonal lattice is an example of
2-colexes where edges are three colorable. b) Corresponding to
each plaquette of the lattice we consider an edge of a hypergraph denoted by yellow (light) color
where the CC can be defined as a quantum CSS state
on that hypergraph. c) Dual of the hypergraph is a triangular
lattice where three vertices of each triangle belong to an edge of
the hypergraph.} \label{col}
\end{figure}
\begin{equation}
B_p =\prod_{i\in p}Z_i ~~~,~~~ A_p =\prod_{i \in p} X_i
\end{equation}
where $i \in p$ refers to all vertices belonging to a plaquette $p$. The quantum CSS state corresponding to the above operators will have the following form:
\begin{equation}
|CC_2 \ra=\prod_{p}(1+A_p ) |0\ra^{\otimes N}.
\end{equation}
It is simple to give an equivalent representation of the above
state on a hypergraph. To this end, we should consider each
plaquette of the lattice as a hyperedge of a hypergraph which
involves all vertices belonging to that plaquette, see
Fig.\ref{col}(b). By such a definition it is simple to find the
dual of this hypergraph. As in Fig.\ref{col}(c), since each vertex
of the $H$ is a member of three hyperedges, the dual hypergraph
will be a triangular lattice. In this way and by the duality
mapping, we conclude that the partition function of a spin model
on triangular lattice with three-body interactions is equal to
inner product of a product state with a CC on the
original trivalent lattice.

Similar to the above argument for 2-colexes, it is simple to find
the duality mapping for 3-colexes. In Fig.\ref{3dcol}, we show a
3-colexes where vertices are four-valent and all edges are colored
by four different colors. A CC on such a 3-colex is
defined by $X$-type and $Z$-type operators in the following form:
\begin{figure}[t]
\centering
\includegraphics[width=7cm,height=4cm,angle=0]{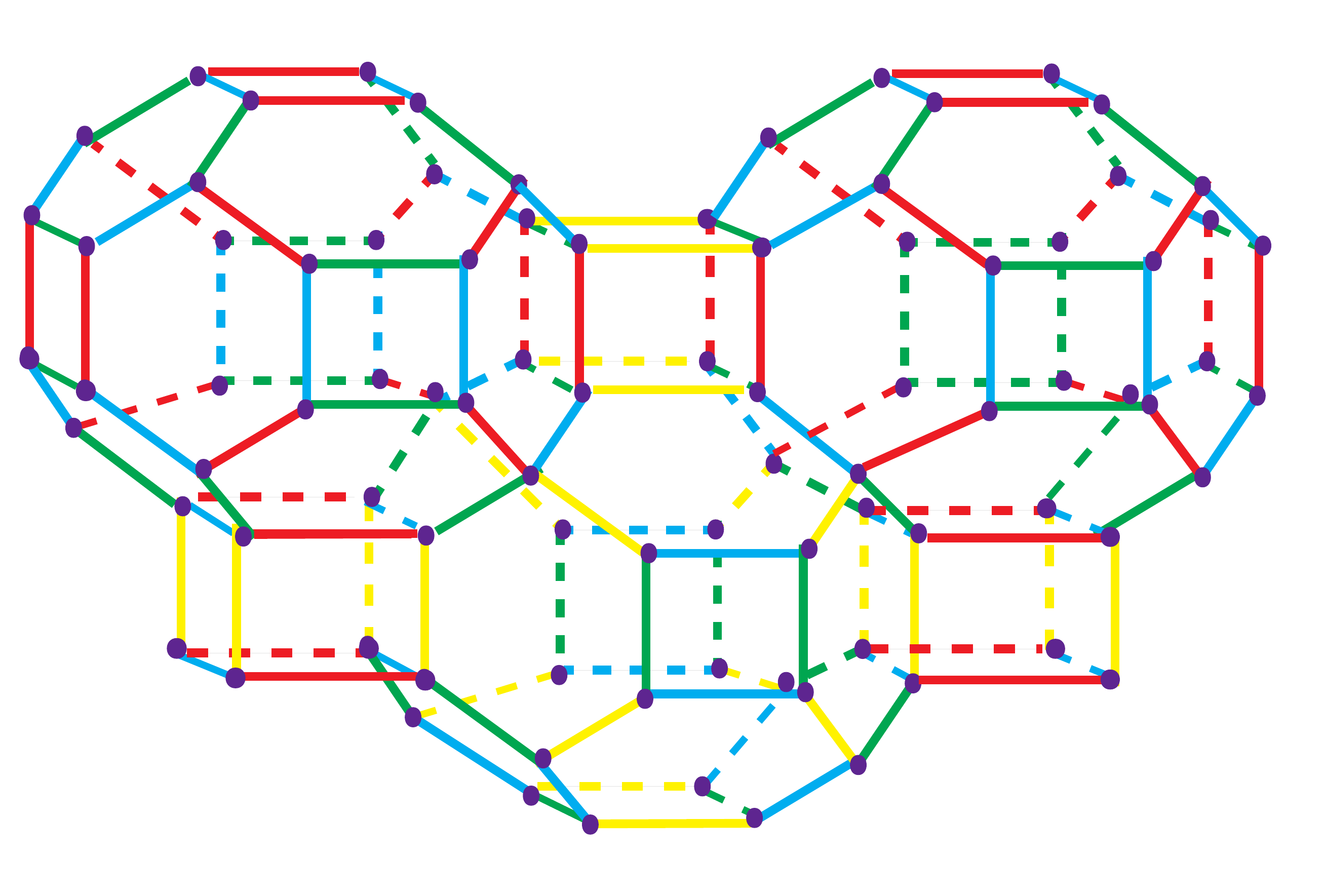}
\caption{(Color online) A 3-colex where all vertices are four valent and all edges are colored by four different colors.} \label{3dcol}
\end{figure}

\begin{equation}
A_c =\prod_{i\in c}X_i ~~~,~~~ B_{f}=\prod_{i\in f}Z_i
\end{equation}

 where $c$ refers to each cell of the lattice and $ f $ refers to
each face of the lattice. Accordingly, the quantum CSS state
corresponding to these operators is in the following form:

\begin{equation}
|CC_3 \ra=\prod_{c}(1+A_c ) |0\ra^{\otimes N}.
\end{equation}
\begin{figure}[t]
\centering
\includegraphics[width=7cm,height=4cm,angle=0]{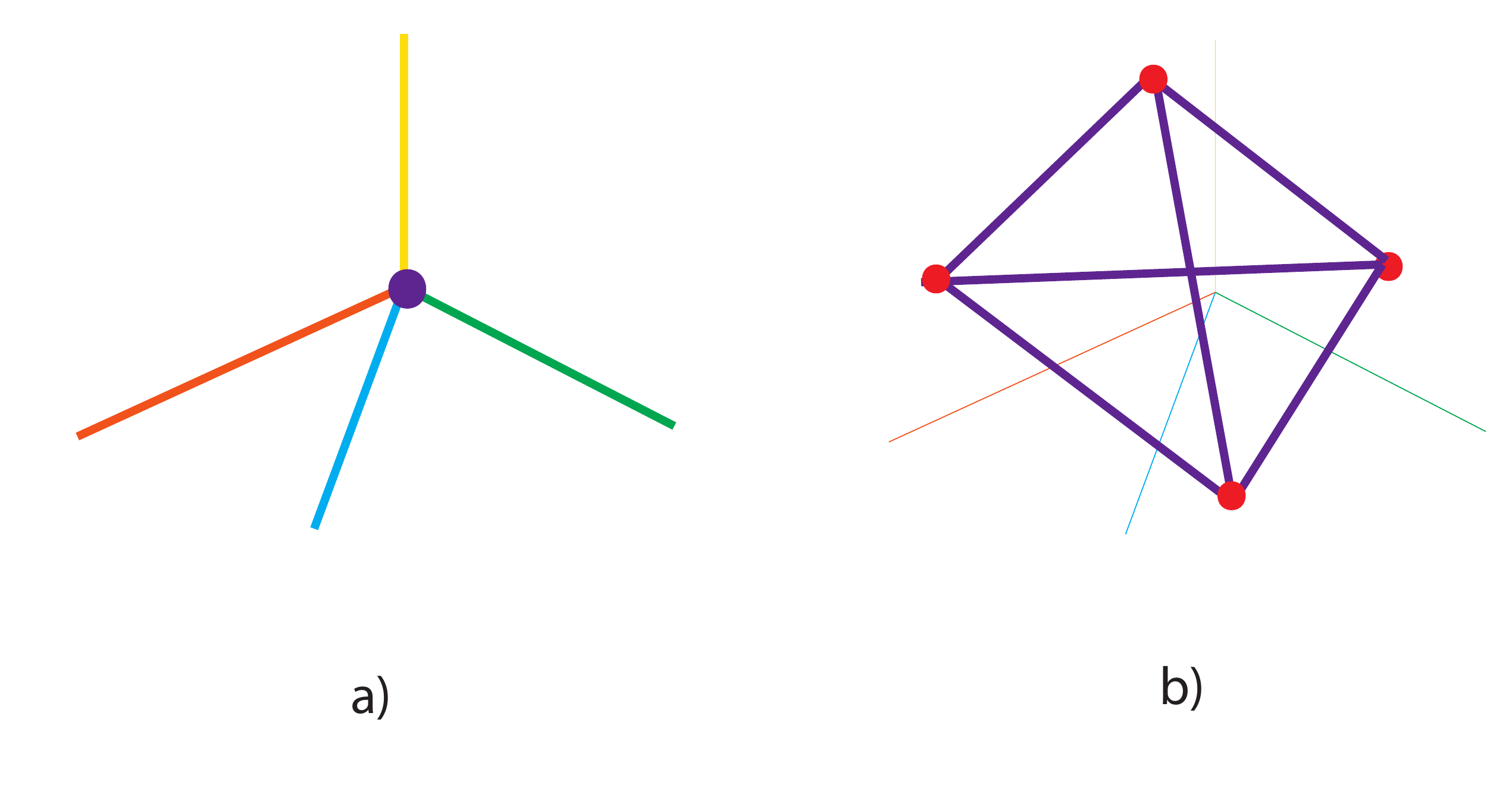}
\caption{(Color online) a) Each vertex of the 3-colex is a four
valent colored by four different colors. b) Since each vertex is a
member of four cells (or edges of the hypergraph) of the 3-colex,
the dual of 3-colex is a tetrahedron lattice.} \label{cc3}
\end{figure}
The hypergraph representation of the above state can easily be
derived by relating cells of the 3-colex to hyperedges of a
hypergraph $H$. Since each vertex of the $H$ is a member of four
hyperedges of the $H$ ( four cells of the 3-colex), the dual
hypergraph will be a tetrahedron lattice with four-colorable
vertices, see Fig.\ref{cc3}. Therefore using our duality mapping,
we have shown that the partition function of a spin model on a
tetrahedron lattice with four-body interactions is related to a
CC on the original 3-colex.

The extension of the above idea to CC in higher dimensions
is straightforward. It is well-known that the dual of a $D$-colex
is a D-simplicial lattice with $(D+1)$-colorable vertices on a
closed $D$-manifold \cite{bombin2015}. We can therefore conclude
that the partition function of a spin model on a D-simplicial
lattice with $(D+1)$-body interactions is related to a CC on a $D$-colex in the following form:
\begin{equation}\label{td1}
\mathcal{Z}_{D-simplicial}=\la \alpha |CC_{D-complex}\ra .
\end{equation}
\section{Critical stability: a case study of duality mapping}\label{s4}
The above mapping between the partition function of a classical
spin model and a quantum CSS state can provide a powerful tool in
order to find how certain well-known property on one side of the
equation would have effects on the other side. This
cross-fertilization could have important consequences.
One clear candidate that can be considered is the non-analytic
property of the classical partition function on the classical
side.  What physics does it correspond to on the quantum side? We
next show that the corresponding CSS state of the critical spin
model has a relative stability to noise, i.e. critical
stability.

In order to define such a concept, suppose that a probabilistic
bit-flip noise is applied to each qubit of the quantum CSS state
with probability $p$. Such a noise can lead to different patterns
of errors that are denoted by $\mathcal{E}$. In other words, each
error $\mathcal{E}$ is a product of Pauli operators $X$ on various
qubits of the CSS state. Suppose that, for a specific error
$\mathcal{E}$, the number of qubits that are affected by Pauli
operators $X$ is equal to $l$. It is clear that the probability of
such an error will be equal to $W_\mathcal{E} =p^l (1-p)^{N-l}$ where $N$ is the number of qubits.
In this way, one can check that $W_\mathcal{E} (p)$ is a
normalized probability where $\sum_{\mathcal{E}}W_\mathcal{E}
(p)=\sum_{l=0}^{N} (\begin{array}{c}
  N \\
  l \\
\end{array})
p^l (1-p)^{N-l} =1 $ .  On the other hand, there is a probability
that pattern of qubits which are affected by the noise is equal to
a stabilizer of CSS state where the CSS state remains in the
stabilizer space. We denote such a probability by $ W_S (p) $
which is defined in the following form:
\begin{equation}
W_S (p)= \sum_{\mathcal{E} \in S}W_\mathcal{E} (p)
 \end{equation}
where $S$ denotes set of stabilizers of the CSS states and $\mathcal{E}\in S $ refers to each error pattern $\mathcal{E}$ which is
equal to one of members of the $S$. Since the stabilizers of
the CSS state do not change the CSS state, we call the above
quantity stability probability. Finally, we define the value of
this quantity as a measure of stability of a CSS state.

Next, we show that the inner product on the right hand of the
duality relation (\ref{sd}) is related to the stability
probability $W_S (p)$. To this end, let us represent the product
state $|\alpha \ra$ in the following form:
\begin{equation}
\frac{1}{2^{\frac{N}{2}}}\prod_{i|v_i \in V}(e^{\beta J}1 + e^{-\beta J}X_{i})|0\ra^{\otimes N},
\end{equation}
which can be again written as:
\begin{equation}\label{sa}
\frac{1}{2^{\frac{N}{2}}[p(1-p)]^{\frac{N}{2}}}\prod_{i|v_i \in V}((1-p)1 + p X_{i})|0\ra^{\otimes N}
\end{equation}
where $N$ is the number of qubits and $p \in [0,\frac{1}{2}]$  is
given by $\frac{p}{1-p}=e^{-2\beta J}$. Consequently, we can write
Eq.(\ref{sd}) as:
\begin{equation}\label{ds}
\mathcal{Z}=\frac{1}{[p(1-p)]^{\frac{N}{2}}}W(p)
\end{equation}
where $W(p)= 2^{M} ~^{N\otimes}\la 0| \prod_{i|v_i \in V}((1-p)1 +
p X_{i})|CSS\ra $. Now, we expand the operator $\prod_{i|v_i \in
V}((1-p)1 + pX_{i})$ in this relation where it will be equal to a
superposition of all errors where the factor of each error term
$\mathcal{E}$ will be equal to $W_{\mathcal{E}}(p)=p^l
(1-p)^{N-l}$. On the other hand, the CSS state in the relation for
$W(p)$ is a superposition of all $X$-type stabilizers of CSS
state. Therefore, the inner product is equal to summation of
$W_\mathcal{E} (p)$'s on all errors that lead to stabilizers of
the CSS state which is equal to $W_S (p)$, or stability
probability of the CSS state i.e. $W(p)=W_S (p)$.

Accordingly, the duality correspondence now finds a new form where
stability probability of a CSS state is related to partition
function of the corresponding classical spin model:
\begin{equation}\label{sdf}
W_S (p)=[p(1-p)]^{\frac{N}{2}}\mathcal{Z}.
\end{equation}
The right hand side goes to zero as $N$ diverges, leading to $W_S
(p)=0$ for finite $\mathcal{Z}$. This seems reasonable, because it
is nearly impossible that bit-flip noise does not lead to an error
in the quantum CSS state \cite{refer}. However, one would have to
reconsider the above, if the classical partition function
$\mathcal{Z}$ also diverges, which could happen at the critical
point of a phase transition.

Under typical situations, $\mathcal{Z}=e^{-\beta F}$ where F is the
Helmholtz free energy. However, near the critical point,
fluctuations become dominant, and one can consider a fluctuation
correction to $\mathcal{Z}=\int_{0}^{\infty} \Omega(E)e^{-\beta
E}dE$, with  $\Omega(E)$ being the density of states. This can
easily be achieved by an expansion about the mean energy of the
system \cite{pathr}, which in leading term is given by
\begin{equation}\label{wq}
\mathcal{Z}=e^{-\beta F}\sqrt{2\pi k T^2 c_v }
\end{equation}
where  $c_v$ is the heat capacity of the classical spin model. The
heat capacity diverges as $(T-T_{cr})^{-\alpha}$ near the critical
phase transition in the thermodynamic limit, with $\alpha$ being a
critical exponent. Therefore,  $\mathcal{Z} \sim
(T-T_{cr})^{-\frac{\alpha}{2}}$. The divergence of the classical
partition function has important ramification for stability of the
corresponding CSS state. Clearly, associated with the critical
temperature $T_{cr}$ there is a critical probability given by
$\frac{p_{cr}}{1-p_{cr}}=e^{-2\beta_{cr} J}$, at which the value
of $W_S (p)$ significantly increases indicating a relative
stability to bit-flip noise at this particular value. We therefore
call this new concept \emph{critical stability} of the CSS state,
defined as the nonzero value of $W_S(p)$ due to critical behavior
of the partition function at a particular noise value $p_{cr}$. We
emphasize that the actual value of $W_S (p)$ does not need to be
large. The system becomes relatively stable at the particular
value of $p=p_{cr}$ since, as $N$ diverges, it is exactly zero
everywhere except at $p_{cr}$. We should emphasize that since
$W_\mathcal{E} (p)$ is a normalized probability function, it will
be clear that $W_S (p)=\sum_{ \mathcal{E} \in S}W_\mathcal{E} (p)$
is always a finite number smaller than 1, since $S$ is a subspace
of the set of all error patterns of $\mathcal{E}$. Specifically,
we emphasize that even at the critical point where partition
function $\mathcal{Z}$ diverges, $W(p)$ remains a finite number
smaller than 1. Furthermore, one can show that critical stability
also exists in the case of phase-flip noise similar to that of
bit-flip noise, see the Appendix.

We would like to emphasize that our concept of critical stability
is very different from the more common concept of robustness in
error correcting threshold for the CSS states which have
previously been considered in the literature. Indeed, since a CSS
state belongs to error correcting codes, it can be protected form
noise by an active error correcting protocol. Specifically, one
can find errors caused by noise by measuring stabilizers of CSS
state. Then it is simple to correct errors by applying suitable
operators \cite{gottesman}. Here, our definition of stability is
completely different and is related to \textit{intrinsic}
stability/robustness of CSS state against bit-flip noise.
Moreover, the usual robustness is a threshold below which the
system can be actively stabilized where critical stability occurs
only at one particular value $p_{cr}$. Furthermore, we note that
the concept of critical stability is relative in a sense that it
is magnified at $p_{cr}$ in relation to other noise probabilities.
The concept of relative stability provides an additional level of
robustness to what one needs in well-know error-correcting
protocols.

The critical behavior of classical spin models is a well-known
phenomenon. However, the concept of critical stability is a new
and interesting property of the quantum CSS state which should
carefully be considered. To this end, we note that the above
concept of critical stability indicates a natural or intrinsic
stability to external noise. On the other hand, a class of CSS
states known as topological CSS states are known to exhibit
stability to other forms of external perturbations. One might
wonder if such stabilities might be related.
Fortunately, the mapping we have provided along with specific
examples discussed in Sec.(\ref{s3}) can help to shed some light
on such a possible relation. We next examine, for some specific
cases, whether such a relation exists.

As was shown in the Sec.(\ref{s3}), each TC on an arbitrary graph
corresponds to the Ising model on the same graph. For example, a
TC on a 1D lattice (GHZ state) maps to a one-dimensional Ising
model which does not exhibit a phase transition, i.e. the TC on a
1D lattice is not stable to noise, $p_{cr}=0$. This is consistent
with the well-known result that there is no topological order in
1D models. However, TC in higher dimensions have topological order
and according to our mapping, they correspond to the Ising model
in higher dimensions which is well-known to exhibit critical
behavior. We also considered classical spin models corresponding
to CC in different dimensions in the previous section. Each CC on
a D-colex (color complex) is mapped to a classical spin model on a
$D$-simplicial lattices that is the dual of the original D-colex.
Although such classical spin models are usually very abstract,
they have a specific symmetry because of their colorability
property. Therefore, one expects a phase transition via a
symmetry-breaking mechanism. For example, a 2D case with
three-body interactions on a triangular lattice with a $Z_2 \times
Z_2$ symmetry has been studied and critical behavior has been
identified \cite{bax1973}, again confirming such a relation.
Another example is provided by the fact that cluster states
correspond to the Ising model in a magnetic field \cite{Nest2007}
which does not exhibit a critical phase transition, again
consistent such a relation, since cluster states do not have a
topological phase. We therefore conjecture that topological CSS
states will exhibit critical stability. In fact, we have not been
able to find any counter-example to our conjecture. If true, it
can provide a practical characterization of topological quantum
CSS states. We again note that the
connection between critical stability and the existence of
topological order seems plausible from a physical point of view
since they both imply stability to external perturbations.

\section{Conclusions}
In this work we provided a duality relation between quantum
CSS states and classical partition function of spin models
(Eq.\ref{sd}). Such duality relation was proved by graph-theoretic
methods and relied heavily on the concept of dual hypergraphs. We
next provided two concrete examples of such a mapping for the
well-known toric code and color code states in various dimensions
in (Eq.\ref{td0}, Eq.\ref{td1}). Using such
correspondence, we introduced the concept of critical stability
(Eq.\ref{sdf}) where it was shown that certain CSS states can
exhibit a natural relative stability to noise at a particular
value of noise probability. Furthermore, we conjectured that
this intrinsic property of CSS states is related to their
topological order. Our general results can open new avenues for
further characterization of quantum CSS states. The generality of
the duality correspondence allows one to look for quantum CSS
states corresponding to well-know classical spin models, or vice
versa. In this way, one might be able to find new complete models
or consider classical simulability of quantum CSS states for MBQC,
just to mention a few possibilities. On the other hand, our
conjecture that critical stability corresponds to topological CSS
states provides a natural characterization of topological states
to external noise. It will be interesting to see if other CSS, as
well as non-CSS, topological states possess critical stability,
thus providing means to characterize topological order in general
quantum states.

\section*{Acknowledgement}
We whould like to thank V. Karimipour and A. T. Rezakhani for their valuable comments on this paper before submitting.

\subsection{APPENDIX: Critical stability against phase-flip noise}
In analogy to critical stability of CSS states against bit-flip
noise, we show that there also exists a critical stability
against a phase-flip noise. To this end, suppose that a Pauli
operator $Z$ is applied to each qubit of a CSS state with
probability $p$. Such a noise leads to an error as a product of
$Z$ operators on various qubits of the CSS state. We denote the
probability of such an error by $V_{\mathcal{E}}(p)$. In analogy to
the bit-flip noise, the above function is also a normalized
probability function. Now, we consider Eq.(12) in a
new form. To this end, we re-write the product state  $|\alpha
\ra$ in the following form:
$$|\alpha\rangle=\frac{1}{2^{\frac{N}{2}}}\prod_{i|v_i \in V}(e^{\beta J}1 + e^{-\beta J}X_{i})|0\ra^{\otimes N}$$
  \begin{equation}
=\prod_{i|v_i \in V}(\cosh(\beta J)1 + \sinh(\beta
J)Z_{i})|+\ra^{\otimes N}.
\end{equation}
Therefore, by a change of variable in the form of $1-2p=e^{-2\beta
J}$ where $p\in [0,\frac{1}{2}]$,  $|\alpha\rangle$ will find the
following form:
\begin{equation}
|\alpha\rangle=\frac{1}{(1-2p)^{\frac{N}{2}}}\prod_{i|v_i \in
V}((1-p)1 + pZ_{i})|+\ra^{\otimes N}.
\end{equation}
The operator $\prod_{i|v_i \in V}((1-p)1 + pZ_{i})$ in the above
relation is equal to superposition of all $Z$-type errors with
corresponding probabilities $V_\mathcal{E} (p)$. On the other
hand, the CSS state can also be written in terms of $Z$-type
stabilizers in the form of $\frac{1}{2^{\frac{N-M}{2}}}
\prod_{\tilde{e}^{*} \in \tilde{E}^{*}}(1+\prod_{m| \tilde{v}_m
\in \tilde{e}^{*}}Z_m ) |+\ra^{\otimes N} $. Here, the operator $
\prod_{\tilde{e}^{*} \in \tilde{E}^{*}}(1+\prod_{m| \tilde{v}_m
\in \tilde{e}^{*}}Z_m )  $ is equal to superposition of all $ Z
$-type stabilizers of the CSS state. In this way, the inner product
of $|\alpha\rangle$ and the CSS state can be interpreted as total
probability that $Z$-type errors lead to stabilizers of the CSS
state. We denote such a probability by $V(p)$ which will be in the
following form:
\begin{equation}
V (p)=\frac{(1-2p)^{\frac{N}{2}}}{2^M}\mathcal{Z}
\end{equation}
Therefore, in the same way as bit-flip noise, the stability
probability due to phase-flip noise will be equal to a product of
quickly decaying function and the partition function of the
corresponding classical spin model, which again leads to critical
stability at the particular value of $p_{cr}$ corresponding to
the critical temperature of the spin model.

It is interesting to note  that the critical probability for
bit-flip noise ($p_{cr}^b$) and phase-flip noise ($p_{cr}^f$) are
different since they are given by
$\frac{p_{cr}^b}{1-p_{cr}^b}=e^{-2\beta_{cr} J}$ and  $1-2p_{cr}^f
=e^{-2\beta_{cr} J}$. In fact,
\begin{equation}\label{q13}
p_{cr}^f=\frac{1}{2} - \frac{p_{cr}^b/2}{(1-p_{cr}^b)},
\end{equation}
which indicates the complimentary nature of critical stability
due to two different noises. That is, if a CSS state exhibits
critical stability at a high value of noise probability for
bit-flip noise it will exhibit such a stability for low values of
probability for phase-flip noise. We emphasize that such critical
probability is not a threshold, and stability occurs at one
particular value of probability and nowhere else. However, one
might wonder if relative stability might occur at the same value
for both types of noises. Clearly, this will happen at
$p_{cr}^b=p_{cr}^f=1-\sqrt{2}/2\approx0.293$. From a more physical
point of view, one would expect this to occur when there is a
symmetry between $X$-type and $Z$-type stabilizers of the given
CSS state.  This is the case with the TC on a square
lattice, for example. In such a model, if we interchange the
$X$-type stabilizers with $Z$-type stabilizers, the model maps to
itself. We therefore expect that the TC on a square
lattice exhibits critical stability at the same value for both
bit-flip and phase-flip noise. Such an expectation is indeed
satisfied by our classical-quantum correspondence as the TC on a square lattice corresponds to the Ising model on a
square lattice which in fact exhibits a critical transition at
$\tanh \beta J =e^{-2\beta J}$ \cite{pathr}. It is easy to check
that this condition is satisfied by Eq.(\ref{q13}) above. We also expect
that the same result would hold for other symmetric CSS states such
as the CC on a hexagonal lattice.

\end{document}